\input harvmac
\Title{\vbox{\baselineskip12pt\hbox{UGVA-DPT 1997/11-992}
\hbox{}}}
{\vbox{\centerline{Surfaces with Long-Range Correlations}
\vskip2pt\centerline{from Non-Critical Strings}}}
\centerline{M. C. Diamantini}
\centerline{Dipartimento di Fisica, Universit\`a di Perugia}
\centerline{via A. Pascoli, I-06100 Perugia, Italy}
\bigskip
\centerline{Carlo A. Trugenberger}
\centerline{D\'epartement de Physique Th\'eorique, Universit\'e de Gen\`eve}
\centerline{24, quai E. Ansermet, CH-1211 Gen\`eve 4, Switzerland}

\vskip .3in
\noindent
We show that the recently proposed confining string theory describes
{\it smooth surfaces} with {\it long-range correlations} for the normal 
components of tangent vectors. These
long-range correlations arise as a consequence of a ``frustrated
antiferromagnetic" interaction whose two main features are non-locality
and a negative stiffness.

\Date{November 1997}

The attempt of solving the quark confinement by means of a
non-critical string theory remains an open and very attractive problem.
In order to cure the well-known problems of the Nambu-Goto action in
4D \ref\gsw{For a review see: M. B. Green, 
J. H. Schwarz and E. Witten, ``Superstring Theory", Cambridge University
Press, Cambridge (1987).} , Polyakov \ref\pola{A. M. Polyakov,
Nucl. Phys. B268 (1986) 406.}\ and Kleinert \ref\klea{H. Kleinert, Phys.
Lett. B174 (1986) 335.} \ proposed a new action with an additional term
proportional to the extrinsic curvature of the world-sheet, the so-called
{\it rigid string}. This term was introduced with a positive sign in order
to avoid crumpling and obtain a smooth world-sheet, a property desired
to match the behaviour of large-$N$ QCD. 

This goal, however, was not attained since the curvature term is infrared
irrelevant \pola ,\klea \ and violent fluctuations lead to the formation
of a finite correlation length for the normals to the surface and to crumpling. 

Recently, Polyakov \ref\polb{A. M. Polyakov, Nucl. Phys. B486 (1997) 23;
see also F. Quevedo and C. A. Trugenberger, Nucl. Phys. B501 (1997) 143.}
\ proposed a new string action to describe the confining phase of gauge
theories. This {\it confining string} theory can be explicitly derived \ref\us{M.
C. Diamantini, F. Quevedo and C. A. Trugenberger, Phys. Lett. B396 (1997)
115.}\ for compact $U(1)$ gauge theories in the phase with a condensate 
of topological defects (magnetic monopoles).

The essential idea of the rigidity term was to obtain smooth surfaces by
enforcing a ``ferromagnetic" alignement of the tangent (or normal) vectors
to the surface. As shown by Polyakov \pola \ and Kleinert \klea , however,
long-range order is prevented by wild fluctuations which cause crumpling.
In this paper we show that {\it long-range correlations} and smooth surfaces 
are indeed obtained for confining strings by a totally different mechanism.
The relevant features of confining strings are two: i) the interaction
between tangent (or normal) vectors to the surface is {\it non-local};
ii) the extrinsic curvature term has a {\it negative} coefficient. 
Note that such a negative stiffness is crucial also to match the
correct high-temperature behaviour of the free-energy of large-$N$ QCD
\ref\klec{H. Kleinert and A. Chervyakov, ``Evidence for
Negative Stiffness of QCD Strings", hep-th/9601030.} .

In our
analogy with spin systems the negative stiffness would correspond to an
``{\it antiferromagnetic}" interaction between tangent (or normal) vectors
to the surface. Non-locality, instead, would mean that this antiferromagnetic
interaction extends beyond ``nearest neighbours". The result is a 
{\it frustrated system}, which can have a very complicated phase structure
\ref\fra{For a review see: E. Fradkin, ``Field Theories of Condensed Matter
Systems", Addison-Wesley, Redwood City (1991).} . It is due to this
non-local, ``frustrated antiferromagnetic" interaction 
that the correlations between
the normal components of tangent vectors are long-range. Moreover, these
correlations display an oscillatory behaviour with period proportional
to $\sqrt{s/T}$ where $s$ is the stiffness and $T$ the renormalized 
string tension.
This is the typical linear dimension of the regions 
on the surface with parallel
orientation of the normals. By adding one half-period to the distances we obtain
anti-parallel orientation and so on.
This is exactly what one would expect for a smooth surface describing
a ``landscape" of hills and valleys of typical area $s/T$. 

The confining string action is induced by a Kalb-Ramond field \ \ref\kr{V. I.
Ogievetsky and V. I. Polubarinov, Sov. J. Nucl. Phys. 4 (1967) 156; M. Kalb
and P. Ramond, Phys. Rev. D9 (1974) 2273.} . In $D$-dimensional Euclidean space 
it is given by
\eqn\csa{\eqalign{{\rm exp} \left(-S_{CS} \right) &= {G\over 
Z\left( B_{\mu \nu} \right)} 
\ \int {\cal D}B_{\mu \nu} \ {\rm exp} \left\{ -S\left( B_{\mu \nu } \right)
+i \int d^D x\ B_{\mu \nu }T_{\mu \nu} \right\} \ ,\cr
T_{\mu \nu} &= {1\over 2} \ \int d^2{\sigma }\ X_{\mu \nu} ({\sigma})
\ \delta ^D ({\bf x}-{\bf x({\sigma})}) \ ,\cr
X_{\mu \nu} &= \epsilon^{ab} {\partial x_{\mu }\over \partial {\sigma }^a}
{\partial x_{\nu}\over \partial {\sigma }^b} \ ,\cr }}
with ${\bf x}({\sigma })$ parametrizing the world-sheet and $G$ a group
factor which will be of no relevance for what follows (we will henceforth
set $G=1$). 
At long distances
the action for the Kalb-Ramond field reduces to
\eqn\akrf{\eqalign{S\left( B_{\mu \nu } \right) &= \int d^D x \ {1\over 12
z^2 \Lambda ^{6-D}}\ H_{\mu \nu \alpha}H_{\mu \nu \alpha} + {1\over 4e^2}
B_{\mu \nu }B_{\mu \nu }\ ,\cr
H_{\mu \nu \alpha } &\equiv \partial_{\mu }B_{\nu \alpha}+\partial_{\nu}
B_{\alpha \mu}+\partial_{\alpha}B_{\mu \nu} \ .\cr}}
It depends on a short-distance cutoff $\Lambda $, a parameter $e$ with
canonical dimension $(4-D)/2$ and one dimensionless
parameter $z$. This action can be explicitly derived (for 
$D$=3 and $D$=4) from a lattice formulation of compact QED in the phase
with condensing magnetic monopoles  \ref\polbook{For a review see:
A. Polyakov, ``Gauge Fields and Strings", Harwood Academic Publishers,
Chur (1987).} . In this case $1/{\Lambda}$ plays the role of the lattice
spacing while $z^2$ is the monopole fugacity; $e$ is the coupling constant
of the underlying gauge theory. Note that, for $z^2\to 0$, only configurations
with $H_{\mu \nu \alpha}=0$ contribute to the partition function: this means
that the Kalb-Ramond field becomes pure gauge, $B_{\mu \nu} = \partial_{\mu}
A_{\nu}-\partial_{\nu}A_{\mu}$, and we recover the partition function of
QED coupled to point-particles described by the boundaries of the original
world-sheet. 

In the continuum formulation above $\Lambda $ can be viewed
as a Higgs mass and, correspondingly, $1/\Lambda$ as a finite thickness of
the string. The mass of the Kalb-Ramond 
field being given by $m=\Lambda ^{3-D/2}z
/e$, the parameter $\tau \equiv \Lambda ^{2-D/2}z/e$ corresponds 
to the ratio (coherence length/penetration
depth) of superconductivity
theory and distinguishes thus between type I ($\tau >1$) and type II ($\tau <1$)
confining strings. This close analogy with superconductivity is not surprising
when one realizes that the same action has been derived for magnetic vortices
in the framework of the Abelian Higgs 
model \ \ref\orl{P.Orland, Nucl. Phys. B428 (1994) 221.} \ .
Note also that an analogous action, although with the opposite 
sign, is relevant to the statistical mechanics of microemulsions
in $D$=3 \ \ref\klez{H. Kleinert, Phys. Lett. 211B (1988) 151.}\ .

In the following we are going to write the confining string action in a form
which lends itself best to analyze the long-distance geometric aspects of
the surfaces it describes. A first step consists in
integrating out the Kalb-Ramond field so that we obtain the action in the
following form:
\eqn\nlf{S_{CS} = {z^2\Lambda ^{6-D}\over 2} \int d^2{\sigma} \int d^2
{\sigma }' \ \sqrt{g({\sigma})} t_{\mu \nu}({\sigma}) \ Y\left( {\bf x}
({\sigma }) - {\bf x}({\sigma}') \right) \ \sqrt{g({\sigma }')}
t_{\mu \nu}({\sigma }') \ ,}
where we have introduced the induced metric
\eqn\inme{\eqalign{g_{ab} &\equiv \partial_a x_{\mu } \partial_b x_{\mu}\ ,\cr
g &\equiv {\rm det}\ g_{ab}\ ,\cr}}
and the tangent tensor
\eqn\tate{t_{\mu \nu } \equiv {1\over \sqrt{2g}}\ X_{\mu \nu }\ .}
The $D$-dimensional Yukawa Green's function $Y$ is given by
\eqn\dygf{Y(r)= N_D\ {1\over (mr)^{D/2-1}} \ K_{D/2-1}(mr)\ ,\qquad \qquad
N_D\equiv {m^{D-2}\over (2\pi)^{D/2}}\ ,}
with $r\equiv |{\bf x}|$ and $K$ a modified Bessel function \ \ref\gr{I. S.
Gradstheyn and I. M. Ryzhik, ``Table of Integrals, Series and Products",
Academic Press, Boston (1980).} \ . In presenting \nlf \ we have left out
additional boundary terms which are of no interest in the present paper.

We proceed by rewriting the non-local kernel in bulk space as a corresponding
non-local kernel on the surface. To this end we introduce a new local coordinate
system around the surface: $\sigma^0, \sigma^1, \chi^2, \dots ,\chi^{D}$.
For $\chi ^i=0$, $i=2,\dots ,D$, the coordinates $\sigma^0$ and $\sigma^1$
describe the original surface; the $\chi^i$, $i=2, \dots, D$, describe
locally flat coordinates orthogonal to the surface. The coordinate
transformation is described by functions 
$x_{\mu }(\sigma^0, \sigma^1, \chi^2, \dots, \chi^D)$
such that $x_{\mu }(\sigma^0, \sigma^1, 0,\dots ,0)$ parametrizes the original
surface. 

The Yukawa Green's function in \nlf \ can now be rewritten as
\eqn\yua{\int d^{D-2}\chi
\int d^{D-2}\chi ' \ \delta^{D-2} (\chi) \ {1\over {m^2-\nabla^2}} \ \delta^D
\left( {\bf x}(\sigma , \chi)-{\bf x}(\sigma ' ,\chi ')\right) 
\ \delta ^{D-2}(\chi ') \ .}
We can now use the transformation rules
\eqn\trru{\eqalign{\delta^D ({\bf x}-{\bf x}') &= {1\over \sqrt{g}}\ \delta^2
(\sigma -\sigma ') \ \delta^{D-2} (\chi - \chi ') \ ,\cr
m^2-\nabla ^2 &= M^2- \nabla^2_{\chi}\ ,\cr
M^2 &\equiv m^2-{\cal D}^2\ ,\cr
{\cal D}^2 &= {\cal D}^a{\cal D}_a= {1\over \sqrt{g}} 
\ \partial_a g^{ab} \sqrt{g} \ \partial_b \ ,\cr }}
where ${\cal D}_a$ denote {\it covariant} derivatives along the surface, to rewrite the
Green's function as
\eqn\yub{\eqalign{Y\left( {\bf x}(\sigma )-
{\bf x}(\sigma ')\right)
&= \int d^{D-2}\chi \ \ \delta^{D-2}(\chi) {1\over {M^2-\nabla^2_{\chi}}}
\delta^{D-2}(\chi ) \ {1\over \sqrt{g}}\ \delta^2(\sigma - \sigma ') \cr
&\equiv {1\over \Lambda ^{4-D}}\ G_D\left( {\tau }, 
\left( {{\cal D}\over \Lambda} \right) ^2 \right) 
\ {1\over
\sqrt{g}}\ \delta^2(\sigma -\sigma ') \ ,\cr}}
where $G_D$ is formally defined via its 
expansion in powers of $({\cal D}/\Lambda)^2$.
In order to compute $G_D$ we first note that
$(1/(M^2-\nabla^2_{\chi})) \delta^{D-2}(\chi )$
is the Yukawa Green's function in $(D-2)$ dimensions and is thus given
by $N_{D-2} (1/(Mr)^{D/2-2}) K_{D/2-2}(Mr)$. The first delta function in
\yub \ tells us then that we have to take this function at $r=0$. For
$D$=3 this gives
\eqn\vdt{G_3\left( {\tau }, \left( {{\cal D}\over \Lambda} \right) ^2 \right)
={1\over 2} \ {1\over \sqrt{\tau ^2 -
\left( {\cal D}\over \Lambda \right) ^2 }} \ .}
For $D\ge 4$, instead, taking the Green's function at $r=0$ leads to an
ultraviolet divergence which requires regularization on the scale $\Lambda $. 
One possibility \orl \   
to regulate this divergence is to write the Yukawa Green's function
at $r=0$ as an integral of $1/(M^2+p^2)$ over Fourier space and to regulate
this integral by a Pauli-Villars procedure. However, this procedure works
only for $D\le 5$ and is thus not well suited for the large $D$ techniques
we shall use later. For completeness, however, we present the result in
the interesting case of $D=4$: 
\eqn\fposs{G^{PV}_4\left( {\tau }, 
\left( {{\cal D}\over \Lambda} \right) ^2 \right) = {1\over 4\pi}\ {\rm ln}
{{\tau ^2 - \left( {\cal D}\over \Lambda \right) ^2 +1 }
\over {\tau ^2 - 
\left( {\cal D} \over \Lambda \right) ^2}}\ .} 
A second possibility, which works for every dimension, 
is to regulate the potential by substituting 
$r\to \sqrt{|\chi|^2+ (1/\Lambda)^2}$ in the Green's function, and only
afterwards taking the limit $|\chi| \to 0$. This leads to
\eqn\sposs{G_D\left( \tau , 
\left( {{\cal D}\over \Lambda}\right) ^2\right)=
{1\over {(2\pi )^{D/2-1}}}\ \sqrt{\tau ^2- 
\left( {{\cal D}\over \Lambda}
\right) ^2} ^{{D\over 2}-2} \ K_{{D\over 2}-2} \left( \sqrt{\tau ^2 
- \left( {{\cal D}\over \Lambda}
\right)^2 } \right) \ .}
Inserting \yub \ into \nlf \ we obtain the confining string action in
the form
\eqn\ist{S_{CS}=\Lambda ^2 \ \int d^2\sigma \sqrt{g} \ \ t_{\mu \nu }
(\sigma ) \ {z^2\over 2}G_D\left( {\tau }, 
\left( {{\cal D}\over \Lambda}\right) ^2\right) \ t_{\mu \nu }
(\sigma ) \ .}

In a last step we use \csa \ and \tate \ to rewrite the action in terms of the
tangent vectors ${\cal D}_a x_{\mu}$ to the surface. To this end we consider
a generic term in the expansion of \ist \ in terms of ${\cal D}^2$:
\eqn\gter{t_{\mu \nu } {\cal D}^{2k} t_{\mu \nu } = {1\over 2}\left(
g^{ac} g^{bd}- g^{ad} g^{bc} \right) \partial_a x_{\mu }\partial_b x_{\nu }
\left( {\cal D}^{2k} \partial _cx_{\mu } \partial _d x_{\nu } +
\partial _c x_{\mu } {\cal D}^{2k}\partial _d x_{\nu } + r_{kcd\mu \nu } 
\right) \ .}
Here $r_{kcd\mu \nu }$ represents the additional terms where the $2k$
covariant derivatives are distributed among $\partial _c x_{\mu }$ and
$\partial _d x_{\nu }$. Using \inme \ we then obtain
\eqn\trans{t_{\mu \nu } \ {\cal D}^{2k}\ t_{\mu \nu } = g^{ab} \ {\cal D}_a
x_{\mu }{\cal D}^{2k} {\cal D}_b x_{\mu } + r_k \ ,}
where $r_0=-1$ and $r_1=0$. Since in this paper we are interested only in
the {\it infrared properties} of confining strings we will neglect the infrared
irrelevant terms $r_k$, $k\ge 2$, and write the confining string action as
\eqn\cofi{\eqalign{S_{CS} &= \Lambda ^2 \int d^2{\sigma} \sqrt{g}\ \ g^{ab}
{\cal D}_a x_{\mu } V_D\left( z, \tau , \left( 
{{\cal D}\over \Lambda } \right) ^2
\right) {\cal D}_b x_{\mu } \ ,\cr 
V_D\left( z, \tau , \left( {{\cal D}\over \Lambda } \right) ^2
\right) &\equiv {z^2\over 2} \left\{ G_D \left( {\tau }, \left( {{\cal D}
\over \Lambda }\right) ^2 \right) - {1\over 2} G_D \left( {\tau }, 0
\right) \right\} \ .}}

At this point it is important to stress which are the crucial aspects of this
action. First of all we point out that the interaction $V_D$ is {\it non-local}.
For a flat background, so that ${\cal D}^2 = \nabla ^2$, it would 
correspond (up to a $\delta $-function) to a potential
\eqn\pot{V_D(\sigma ) = {z^2 \Lambda ^2 \over {2 (2\pi )^{D/2}}} 
\ \left( {1\over \tau}
\sqrt{1+ \left( \Lambda |\sigma | \right) ^2 } \right) ^{1-D/2} 
\ K_{D/2-1} \left( {\tau } \sqrt{\tau + \left( \Lambda |\sigma | \right) ^2 }
\right) \ ,}
which is positive for all $|\sigma |$. However, the exact form of this
potential at short distances is irrelevant for the infrared properties of
confining strings. Only the first two terms in the expansion of $V_D$,
\eqn\expa{\eqalign{V_D &\left( z, \tau , 
\left( {{\cal D}\over \Lambda} \right) ^2 \right)
= t_0 + s \left( {{\cal D}\over \Lambda } \right) ^2 + 
w \left( {{\cal D}\over \Lambda } \right) ^4 +\dots \ ,\cr
t_0 &= {z^2\over {4 (2\pi )^{D/2-1}}} \ \tau ^{D/2-2}\ K_{D/2-2}(\tau ) \ ,\cr
s &={z^2\over {4 (2\pi )^{D/2-1}}}\ \tau ^{D/2-3} \ K_{D/2-3} (\tau )\ ,\cr
w &={z^2\over {16 (2\pi )^{D/2-1}}}\ \tau ^{D/2-4} \ K_{D/2-4} (\tau )\ ,\cr }}
will be relevant. Nevertheless, the presence of the remaining terms is crucial
to guarantee a good ultraviolet behaviour of $V_D$; for example the absence
of zeros at finite momentum in the Fourier transform of $V_D$.

Note the sign of the second term in \expa , which shows that confining strings
have {\it negative stiffness}. A non-local action with negative stiffness, 
analogous to \cofi , has also been recently proposed in \klec , where it was
found that a negative stiffness is crucial in order to match the correct 
high-temperature behaviour of large-$N$ QCD.

In order to expose clearly the mechanism at work in confining strings we
compare \cofi \ with the action of the rigid string \pola \ \klea , which
has
\eqn\rigid{V_D^{\rm rigid} = {\mu _0 \over 2\Lambda ^2} - {1\over \alpha}
\left( {{\cal D}\over \Lambda } \right) ^2 \ ,}
with $\mu _0$ the bare string tension. The basic idea of rigid strings was
to obtain smooth surfaces by enforcing the alignement of the tangent
vectors ${\cal D}_a x_{\mu }$ with the ``ferromagnetic" interaction embodied
by the positive stiffness, i.e. the second term in \rigid . However, violent
fluctuations contrast this mechanism, generate a finite correlation length and
prevent long-range order. 

In confining strings, the interaction is ``{\it antiferromagnetic}", as can
be readily recognized by the positivity of the potential \pot \ or by the
negative stiffness in \expa . However, this ``antiferromagnetic" interaction
is non-local, it goes beyond ``nearest neighbours" in the analogy with spin
systems. This creates {\it frustration} \fra \ since nearby tangent vectors
have to find a compromise between parallel and anti-parallel alignement.
In the following we shall compute the correlation functions arising
from this mechanism, which is totally different from the one at work in the
rigid string.

To this end we shall use standard large $D$ techniques \ref\dav{H. Kleinert,
Phys. Rev. Lett. 58 (1987) 1915; P. Olesen and S. K. Yang, Nucl. Phys.
B283 (1987) 73; E. Braaten, R. D. Pisarski and S. M. Tse, Phys. Rev. Lett.
58 (1987) 93; F. David and E. Guitter, Nucl. Phys. B295 (1988) 332.}
\ starting from the action \cofi . We first introduce a (dimensionless) 
Lagrange multiplier
to enforce the constraint $g_{ab}=\partial _ax_{\mu }\partial_bx_{\nu }$,
\eqn\lamu{S_{CS} \to S_{CS}+ \Lambda ^2 \int d^2\sigma \sqrt{g} \ \ \lambda
^{ab} \left( \partial _a x_{\mu } \partial _bx_{\nu } - g_{ab} \right) \ .}
We then parametrize the world-sheet in the Gauss map as
\eqn\gauss{x_{\mu } (\sigma ) = \left( \sigma _0, \sigma _1, \phi ^i (\sigma )
\right) \ ,\qquad \qquad i=2, \dots , D \ ,}
where $-\beta /2\le \sigma_0 \le \beta /2$, $-R/2 \le \sigma ^1 \le R/2$ and 
$\phi ^i(\sigma )$ describe the transverse fluctuations. With the usual
isotropy Ansatz
\eqn\iso{g_{ab}=\rho \ \delta_{ab} \ ,\qquad \qquad \lambda ^{ab} =
\lambda \ g^{ab} \ ,}
for the metric and the Lagrange multiplier of infinite systems 
($\beta ,R \to \infty $) at the saddle
point we obtain
\eqn\gmapac{S_{CS}=2\Lambda ^2 \int d^2\sigma \ \left( t_0+\lambda
(1-\rho ) \right) + \Lambda ^2 \int d^2\sigma \ \partial_a\phi ^i
\left( \lambda + V_D\left( z, \tau , \left( {{\cal D}\over \Lambda }\right) ^2
\right) \right) \ \partial_a \phi ^i\ .}

Integrating over the transverse fluctuations $\phi ^i$ produces a Tr Log
with a factor $D-2$ which suppresses the fluctuations of $\lambda $ and
$\rho $ for large $D$. These are thus determined by the two saddle-point
equations
\eqn\saddle{\eqalign{{{\rho-1}\over \rho} &= {{D-2}\over{16 \pi ^2}}
\ \int _{\rm reg} d^2p\ 
{1\over {\lambda + V_D \left( z, \tau, p^2\right) }}\ ,\cr
\lambda &= {{D-2}\over {16\pi ^2}}\ \int_{\rm reg}d^2p \ {\rm ln}
\left\{ p^2 \left( \lambda + V_D 
\left( z, \tau, p^2\right) \right) \right\} \ ,\cr }}
where the subscript ``reg" means that the integrals are ultraviolet regulated
on the scale 1 (note that $p$ above is 
the intrinsic momentum in units of $\Lambda $).
The details of the regularization procedure used are of no importance in
what follows. In the infinite area limit
the effective action is given by 
\eqn\efac{S_{CS}^{\rm eff}= 2\Lambda ^2 \ \left( t_0+\lambda \right)
\ A_{\rm ext} \ ,}
where $A_{\rm ext}=\beta R$ is the extrinsic, physical area. 
From here
we read off the renormalized string tension
\eqn\tension{\eqalign{T &= 2\Lambda ^2 \ t\ ,\cr
t &\equiv t_0+\lambda \ .\cr }}

We postpone a detailed analysis of the saddle-point equations \saddle \ to
a forthcoming publication. Here we shall only address the main points.
First of all
let us remark that in the deep type I regime $\tau \gg 1$
the confining string reduces to the Nambu-Goto string since $V_D\left( z,
\tau, p^2 \right) $ is essentially constant for $0<p\le 1$, as is clear
from \sposs \ and \cofi . Note also that, for finite $\tau $, 
infrared finiteness prevents the
vanishing of the string tension, as can be readily recognized by using the
expansion \expa \ in the first of the two saddle-point equations \saddle .
This is analogous to the rigid string and means that the confining string
must be viewed as an effective string theory with a fundamental microscopic
length scale. However, contrary to the case of the rigid string, this
does not prevent long-range correlations, as we show below. 

Very interesting is also the deep type II regime in which $\tau \to 0$.
In this case the mass of the original Kalb-Ramond field in \akrf \ vanishes
and the interaction $Y\left( {\bf x}(\sigma )- {\bf x}(\sigma ')\right)$
in \nlf \ becomes a long-range {\it Coulomb} potential. For $\tau \to 0$
the expansion \expa \ is no more valid, as can be seen from the divergence
of the coefficient of $\left( {\cal D}/\Lambda \right) ^{2k}$ for
sufficiently high $k$. In this case one has to keep the full expression
\eqn\full{V_D\left( z, 0, p\right) = {z^2\over {2 (2\pi )^{D/2-1}}}
\ \left\{ p^{D/2-2} K_{D/2-2} (p) -{1\over 2} \ {\rm lim}_{p\to 0} 
\ p^{D/2-2} K_{D/2-2}(p) \right\} \ .}
For sufficiently large $D$ the expansion
\eqn\large{V_D( z, 0, p) = t_0 - sp^2 + \dots \ ,}
is still valid to this order. This expansion, however, breaks down for
$D\le 6$. In particular, for $D<6$ the infrared finiteness of the first
saddle point equation does not prevent anymore a solution with
vanishing string tension, as is best exemplified by the case $D=5$, for
which
\eqn\casefive{V_5( z, 0, p )= {z^2\over 16 \pi}- {z^2\over 8 \pi } p+\dots \ ,}
so that the denominator in the saddle point equation becomes
$t-\left( z^2/8\pi \right) p+\dots $.

In the following we shall assume that we are in the {\it string phase}, i.e.
that solutions exist for both saddle-point equations (note that we know
by construction that at least another, gauge phase exists for $z\to 0$).
In the string phase we would like to compute the correlation function
\eqn\corr{g_{ab}(\sigma -\sigma ') \equiv \langle \partial_a \phi ^i (\sigma )
\ \partial_b \phi ^i (\sigma ') \rangle \ ,}
which describes the scalar product of the components of the tangent vectors
normal to the reference plane $\left( \sigma ^0, \sigma ^1 \right) $ at
different points on the surface. This is obtained from \gmapac \ as
\eqn\corrb{g_{ab} \left( \sigma - \sigma ' \right) = {1\over (2\pi )^2}
\ \int d^2p \ {\delta _{ab}\over 
{\lambda + V_D\left( z, \tau ,p^2\right) }}
\ \ {\rm e}^{i\Lambda \sqrt{\rho } p (\sigma -\sigma ')} \ ,}
and reduces for large $D$ to
\eqn\new{g_{ab} \left( \sigma - \sigma ' \right) = {1\over (2\pi )^2}
\ \int d^2p \ {\delta_{ab}\over {t-sp^2+\dots }} 
\ \ {\rm e}^{i\Lambda \sqrt{\rho } p (\sigma -\sigma ')}\ .}
For $s/t <1$ we are in the Nambu-Goto limit and the correlation reduces
essentially to a $\delta $-function. More interesting is the opposite
regime $s/t \gg 1$ in which the negative stiffness has a crucial influence.

The dominant large-distance behaviour of the correlation function \new \ can 
be derived independently of the ultraviolet details of the interaction: 
\eqn\resu{\eqalign{g_{ab}(\sigma -\sigma ') &\simeq 
N \ \delta_{ab} \ {1\over \sqrt{|\sigma -\sigma '|}} 
\ {\rm sin} \left( \sqrt{{t\rho\over s}} \Lambda 
|\sigma -\sigma '| - {\pi\over 4}\right) \ ,\cr
N &= {1\over 4} \ \left( 
s^3t\Lambda ^2 \rho \right) ^{{-1\over 4}}\ .\cr}}
However, once derived, it is easiest to convince oneself of the
correctness of this result by computing backwards its Fourier transform.
To this end we first point out 
that \resu \ represents the asymptotic behaviour of the
Neumann function 
$(1/4s) N\left( \sqrt{t\rho /s} \Lambda |\sigma -\sigma '|
\right) $. Computing the two-dimensional 
Fourier transform of this function one finds
\gr \ $1/(t-sp^2)$ for all $p<\sqrt{t/s}$ i.e. one reproduces exactly
the infrared behaviour of the correlation \new \ in momentum space.

The correlation function \resu \ is {\it long-range} in the usual sense
that $\int d^2\sigma \ g_{aa}(\sigma ) $ is infrared divergent. Notice the
oscillatory behaviour: it tells us that the normal components of the
tangent vectors have parallel alignement 
in regions of intrinsic area $2\pi ^2 s/T$.
By adding to the intrinsic distances one half-period 
$\sqrt{2\pi ^2 s/T}$ we get
anti-parallel alignement and so on. The normal components of the tangent vectors
describe how the surface is ``growing out" of the reference plane
$\left( \sigma ^0, \sigma ^1 \right) $. The 
correlations \resu \ indicate thus that the surface oscillates in and out
of the reference plane with a typical intrinsic length scale $\sqrt{2\pi ^2
s/T}$. This oscillation length scale is determined by the ratio of the
``antiferromagnetic" coupling $s$ and the renormalized string tension $T$,
which contains via $\lambda $ all the effects of the non-local interaction,
i.e. all the frustration effects. For $s/t \gg 1$ the oscillation scale
is much larger than the microscopic length scale $1/\Lambda $ and we have
a smooth surface. As a concluding remark we record the ratio $s/t$ for
$D=6$:
\eqn\sovert{{s\over t} = {z^2\over 16 \pi ^2} {K_0(\tau )\over t}\ .}
If the (dimensionless) string tension $t$ does not diverge, this ratio and
thus the size of the alignement regions diverges in the Coulomb limit
$\tau \to 0$.

\listrefs
\end